\documentclass[a4paper,12pt]{article}
\usepackage[english]{babel}
\usepackage{graphicx,color}
\usepackage{amsmath,amsthm,amssymb,amsfonts}
\usepackage{epsfig}
\usepackage[T1]{fontenc}
\usepackage[cp1250]{inputenc}
\usepackage[bookmarks,colorlinks]{hyperref}
\usepackage{fancyhdr}
\fancypagestyle{plain}{%
\fancyhead{}

}

\hypersetup{colorlinks,%
linkcolor=magenta,%
urlcolor=blue,%
citecolor=red}

\title{CP violation, massive neutrinos, and its chiral condensate: new results from Snyder noncommutative geometry}

\author{
\textbf{\L ukasz Andrzej Glinka}\vspace*{15pt}\\
\href{mailto:laglinka@gmail.com}{\tt{laglinka@gmail.com}}\vspace*{15pt}\\
\emph{International Institute for Applicable}\\
\emph{Mathematics \& Information Sciences,}\\
\emph{Hyderabad (India) \& Udine (Italy),}\vspace*{10pt}\\
\emph{B.M. Birla Science Centre,}\\
\emph{Adarsh Nagar, 500 063 Hyderabad, India}
}
\date{\today}

\begin{document}

\maketitle
\begin{abstract}
The Snyder model of a noncommutative geometry due to a minimal scale $\ell$, \emph{e.g.} the Planck or the Compton scale, yields $\ell^2$-shift within the Einstein Hamiltonian constraint, and $\gamma^5$-term in the free Dirac equation violating CP symmetry manifestly.

In this paper the Dirac equation is reconsidered. In fact, there is no any reasonable cause for modification of the Minkowski hyperbolic geometry of a momentum space. It is the consistency -- in physics phase space, spacetime (coordinates), and momentum space (dynamics) are independent mathematical structures. It is shown that the modified Dirac equation yields the kinetic mass generation mechanism for the left- and right-handed Weyl chiral fields, and realizes the idea of neutrinos receiving mass due to CP violation. It is shown that the model is equivalent to the gauge field theory of composed two 2-flavor massive fields. The global chiral symmetry spontaneously broken into the isospin group leads to the chiral condensate of massive neutrinos. This result is beyond the Standard Model, but in general can be included into the theory of elementary particles and fundamental interactions.
\end{abstract}
\newpage

\section{Introduction}

In 1947 the American physicist H. S. Snyder, for elimination of the infrared catastrophe in the Compton effect, proposed employing the model \cite{snyder}
\begin{equation}\label{nd}
\dfrac{i}{\hbar}[x,p]=1+\alpha\left(\dfrac{\ell}{\hbar}\right)^2p^2\quad,\quad \dfrac{i}{\hbar}[x,y]=O(\ell^2)\quad,
\end{equation}
with $p$ - a particle's momentum, $x$, $y$ - space points, $\ell$ - a fundamental length scale, $\hbar$ - the Planck constant, $\alpha\sim1$ - a dimensionless constant, $[\cdot,\cdot]$ - an appropriate Lie bracket. For the Lorentz and Poincar\'{e} invariance modified due to $\ell$, Snyder considered a momentum space of constant curvature isometry group, \emph{i.e.} the Poincar\'{e} algebra deformation into the De Sitter one.

The model (\ref{nd}) is a noncommutative geometry and a deformation (Basics and applications: \emph{e.g.} Ref. \cite{ng}). Let us see first it in some general detail. Let $A$ - an associative Lie algebra, $\tilde{A}=A[[\lambda]]$ - the module due to the ring of formal series $\mathbb{K}[[\lambda]]$ in a parameter $\lambda$. A deformation of $A$ is a $\mathbb{K}[[\lambda]]$-algebra $\tilde{A}$ such that $\tilde{A}/\lambda\tilde{A}\approx A$. If $A$ is endowed with a locally convex topology with continuous laws, \emph{i.e.} a topological algebra, then $\tilde{A}$ is topologically free. If in $A$ composition law is ordinary product and related bracket is $[\cdot, \cdot]$, then $\tilde{A}$ is associative Lie algebra if for $f,g\in A$ a new product $\star$ and bracket $[\cdot, \cdot]_\star$ are
\begin{eqnarray}
f\star g&=&fg+\sum_{n=1}^\infty\lambda^nC_n(f,g),\label{star}\\
\left[f,g\right]_\star&\equiv&f\star g-g\star f=\left[f,g\right]+\sum_{n=1}^\infty\lambda^nB_n(f,g),\label{starb}
\end{eqnarray}
where $C_n$, $B_n$ are the Hochschild and Chevalley 2-cochains, and for $f,g,h\in A$ hold
$(f\star g)\star h = f\star(g\star h)$ and $[[f,g]_\star,h]_\star+[[h,f]_\star,g]_\star+[[g,h]_\star,f]_\star=0$. For each $n$ and $j,k\geqslant1$, $j+k=n$ the equations are satisfied
\begin{eqnarray}
\!\!\!\!\!\!\!\!bC_n(f,g,h)\!\!\!&=&\!\!\!\!\sum_{j,k}\left[C_j\left(C_k(f,g),h\right)-C_j\left(f,C_k(g,h)\right)\right],\\
\!\!\!\!\!\!\!\!\partial B_n(f,g,h)\!\!\!&=&\!\!\!\!\sum_{j,k}\left[B_j\left(B_k(f,g),h\right)+B_j\left(B_k(h,f),g\right)+B_j\left(B_k(g,h),f\right)\right],
\end{eqnarray}
where $b$, $\partial$ are the Hochschild and Chevalley coboundary operators - $b^2=0$, $\partial^2=0$. Let $C^\infty(M)$ - an algebra of smooth functions on a differentiable manifold $M$. Associativity yields the Hochschild cohomologies. An antisymmetric contravariant 2-tensor $\theta$ trivializing the Schouten–Nijenhuis bracket $[\theta,\theta]_{SN}=0$ on $M$, defines the Poisson bracket $\{f,g\}=i\theta df\wedge dg$ with the Jacobi identity and the Leibniz rule; $(M,\{\cdot,\cdot\})$ is called a Poisson manifold.

In 1997 the Russian mathematician M. L. Kontsevich \cite{kontsevich} defined deformation quantization of a general Poisson differentiable manifold. Let  $\mathbb{R}^d$ endowed with a Poisson bracket $\alpha(f,g)=\sum_{1\leqslant i,j\eqslantless n}\alpha^{ij}\partial_if\partial_jg$, $\partial_k=\partial/\partial x^k$, $1\leqslant k\leqslant d$. For $\star$-product, $n\geqslant0$, exists a family $G_{n,2}$ of $(n(n+1))^n$ oriented graphs $\Gamma$. $V_\Gamma$ - the set of vertices of $\Gamma$; has $n+2$ elements - 1st type $\{1,\ldots,n\}$, 2nd type $\{\bar{1},\bar{2}\}$. $E_\Gamma$ - the set of oriented edges of $\Gamma$; has $2n$ elements. There is no edge starting at a 2nd type vertex. Star(k) - $E_\Gamma$ starting at a 1st type vertex $k$ with cardinality $\sharp k=2$, $\sum_{1\leqslant k \leqslant n}\sharp k = 2n$. $\{e^1_k,\ldots,e^{\sharp k}_k\}$ are the edges of $\Gamma$ starting at vertex $k$. Vortices starting and ending in the edge $v$ are $v = (s(v),e(v))$, $s(v)\in\{1,\ldots,n\}$ and $e(v)\in\{1,\ldots,n;\bar{1},\bar{2}\}$. $\Gamma$ has no loop and no parallel multiple edges. A bidifferential operator $(f,g)\mapsto B_\Gamma(f,g)$, $f,g\in C^\infty(\mathbb{R}^d)$ is associated to $\Gamma$. $\alpha^{e^1_ke^2_k}$ are associated to each 1st type vertex $k$ from where the edges $\{e^1_k,e^2_k\}$ start; $f$ is the vertex $1$, $g$ is the vertex $\bar{2}$. Edge $e^1_k$ acts $\partial/\partial x^{e^1_k}$ on its ending vertex. $B_\Gamma$ is a sum over all maps $I:E_\Gamma\rightarrow\{1,\ldots,d\}$
\begin{equation}
\!\!\!B_\Gamma(f,g)=\sum_I\left(\prod_{k=1}^n\prod_{k'=1}^n\partial_{I(k',k)}\alpha^{I(e^1_k)I(e^2_k)}\right)\!\!\left(\prod_{k_1=1}^n\partial_{I(k_1,\bar{1})}f\right)\!\!\left(\prod_{k_2=1}^n\partial_{I(k_2,\bar{2})}g\right).\!\!\!
\end{equation}
Let $\mathcal{H}_n$ - an open submanifold of $\mathbb{C}^n$, the configuration space of $n$ distinct points in $\mathcal{H}=\{x\in\mathbb{C}|\Im(z)>0\}$ with the Lobachevsky hyperbolic metric. For the vertex $k$, $1\leqslant k \leqslant n$, $z_k\in\mathcal{H}$ - a variable associated to $\Gamma$. The vertex $1$ associated to $0\in\mathbb{R}$, the vertex $\bar{2}$ to $1\in\mathbb{R}$. If $\tilde{\phi}_v=\phi(s(v),e(v))$ - a function on $\mathcal{H}_n$, associated to $v$, with $\phi:\mathcal{H}_2\rightarrow\mathbb{R}/2\pi\mathbb{Z}$ - the angle function
\begin{equation}
  \phi(z_1,z_2)=\mathrm{Arg}\dfrac{z_2-z_1}{z_2-\bar{z}_1}=\dfrac{1}{2i}\mathrm{Log}\dfrac{\bar{z}_2-z_1}{z_2-\bar{z}_1}\dfrac{z_2-z_1}{\bar{z}_2-\bar{z}_1},
\end{equation}
then $w(\Gamma)\in\mathbb{R}$, the integral of $2n$-form, is a weight associated to $\Gamma\in G_{n,2}$
\begin{eqnarray}\label{weight}
w(\Gamma)=\dfrac{1}{n!(2\pi)^{2n}}\int_{\mathcal{H}_n}\bigwedge_{1\leqslant k \leqslant n}\left(d\tilde{\phi}_{e^1_k}\wedge d\tilde{\phi}_{e^2_k}\right).
\end{eqnarray}
The weight does not depend on the Poisson structure or the dimension $d$. On $(\mathbb{R}^d,\alpha)$ the Kontsevich $\star$-product maps $C^\infty(\mathbb{R})\times C^\infty(\mathbb{R})\rightarrow C^\infty(\mathbb{R})[[\lambda]]$
\begin{equation}\label{konts}
  (f,g)\mapsto f\star g=\sum_{n\geqslant 0}\lambda^nC_n(f,g)\quad,\quad  C_n(f,g)=\sum_{\Gamma\in G_{n,2}}w(\Gamma)B_\Gamma(f,g),
\end{equation}
with $C_0(f,g)=fg$, $C_1(f,g)=\{f,g\}_\alpha=\alpha df\wedge dg$. Equivalence classes of (\ref{konts}) are bijective to the Poisson brackets $\alpha_\lambda=\sum_{k\geqslant0}\lambda^k\alpha_k$ ones. For linear Poisson structures, \emph{i.e.} on coalgebra $A^\star$, (\ref{weight}) of all even wheel graphs vanishes, and (\ref{konts}) coincides with the $\star$-product given by the Duflo isomorphism. This case allows to quantize the class of quadratic Poisson brackets that are in the image of the Drinfeld map which associates a quadratic to a linear bracket.

Let us consider the deformations of phase-space and space given by the parameters $\lambda_{ph}$, $\lambda_s$ being
\begin{equation}
\lambda_{ph}=\dfrac{\alpha i\hbar}{2}\quad,\quad\lambda_s=\dfrac{i\beta}{2}\quad,\quad\alpha\sim1,
\end{equation}
and leading to the star products (\ref{star}), or equivalently the Kontsevich ones (\ref{konts}), on the phase space $(x,p)$ and between two distinct space points $x$ and $y$
\begin{eqnarray}
x\star p&=&px+\sum_{n=1}^\infty \left(\dfrac{\alpha i\hbar}{2}\right)^nC_n(x,p),\label{star1}\\
x\star y&=&xy+\sum_{n=1}^\infty \left(\dfrac{i\beta}{2}\right)^nC_n(x,y),\label{star2}
\end{eqnarray}
where $C_n(x,p)$, $C_n(x,y)$ are the appropriate Hochschild cochains in (\ref{konts}). The brackets arising from the star products (\ref{star1}) and (\ref{star2}) are
\begin{eqnarray}
\left[x,p\right]_\star&=&\left[x,p\right]+\sum_{n=1}^\infty \left(\dfrac{\alpha i\hbar}{2}\right)^nB_n(x,p),\label{starb1}\\
\left[x,y\right]_\star&=&\left[x,y\right]+\sum_{n=1}^\infty \left(\dfrac{i\beta}{2}\right)^nB_n(x,y),\label{starb2}
\end{eqnarray}
where $B_n(x,p)$, $B_n(x,y)$ are the Chevalley cochains. By using $[x,p]=-i\hbar$ and $[x,y]=0$, and taking the first approximation of (\ref{starb1}) and (\ref{starb2}) one obtains
\begin{eqnarray}
\left[x,p\right]_\star=-i\hbar+\dfrac{\alpha i\hbar}{2}B_1(x,p)\quad,\quad\left[x,y\right]_\star=\dfrac{i\beta}{2}B_1(x,y).\label{starb1a}
\end{eqnarray}
or in the Dirac ''method of classical analogy'' form \cite{dirac}
\begin{eqnarray}
\dfrac{1}{i\hbar}\left[p,x\right]_\star=1-\dfrac{\alpha}{2}B_1(x,p)\quad,\quad\dfrac{1}{i\hbar}\left[x,y\right]_\star=\dfrac{\beta}{2\hbar}B_1(x,y).\label{starb2a}
\end{eqnarray}
Because, however, for $f,g\in C^\infty(M)$: $B_1(f,g)=2\theta(df\wedge dg)$, so one has
\begin{eqnarray}
\dfrac{1}{i\hbar}\left[p,x\right]_\star=1-\dfrac{\alpha}{\hbar}(dx\wedge dp)\quad,\quad\dfrac{1}{i\hbar}\left[x,y\right]_\star=\dfrac{\beta}{\hbar} dx\wedge dy,\label{starb2b}
\end{eqnarray}
where $\hbar$ in first relation was introduced for dimensional correctness. Taking into account the simplest space lattice with a fundamental length scale $\ell$
\begin{equation}
x=ndx\quad,\quad d x=\ell\quad,\quad n\in\mathbb{Z}\quad\longrightarrow\quad\ell=\dfrac{l_0}{n}e^{1/n}\quad,\quad\lim_{n\rightarrow\infty}\ell=0,
\end{equation}
where $l_0>0$ is a constant, and the De Broglie coordinate-momentum relation
\begin{equation}
  p=\dfrac{\hbar}{x}
\end{equation}
one receives finally the brackets
\begin{eqnarray}
\dfrac{i}{\hbar}\left[x,p\right]_\star=1+\dfrac{\alpha}{\hbar^2}\ell^2p^2\quad,\quad\dfrac{i}{\hbar}\left[x,y\right]_\star=-\dfrac{\beta}{\hbar}\ell^2,\label{nd1}
\end{eqnarray}
that are defining the Snyder model (\ref{nd}).

In the 1960s the Soviet physicist M. A. Markov \cite{markov} proposed to take a fundamental length scale as the Planck length $\ell=\ell_{Pl}=\sqrt{\strut{\dfrac{\hbar c}{G}}}$, and suppose that a mass $m$ of any elementary particle is $m\leqslant M_{Pl}=\dfrac{\hbar}{c\ell_{Pl}}=\sqrt{\strut{\dfrac{G\hbar}{c^3}}}$. Using this idea, since 1978 the Soviet-Russian theoretician V. G. Kadyshevsky and collaborators (See \emph{e.g.} papers in Ref. \cite{kadyshevsky}) have studied widely some aspects of the Snyder noncommutative geometry model. Recently also V. N. Rodionov has developed the Kadyshevsky current independently \cite{radionov}. The problems discussed in this paper seem to be more related to a general current \cite{book}, where the Snyder model (\ref{nd}) is partially employed.

Beginning 2000 the Indian scholar B. G. Sidharth \cite{sidharth} showed that in spite of self-evident Lorentz invariance of the structural deformation (\ref{nd}), in general the Snyder modification both breaks the Einstein special equivalence principle as well as violates the Lorentz symmetry so celebrated in relativistic physics. In that case the Einstein Hamiltonian constraint receives an additional term proportional to 4th power of spatial momentum of a relativistic particle and 2nd power of $\ell$ that is a minimal scale, \emph{e.g.} the Planck scale or the Compton one, of a theory (Cf. Ref. \cite{bgs2008foop1})
\begin{equation}\label{ss}
  E^2=m^2c^4+c^2p^2+\alpha\left(\dfrac{c}{\hbar}\right)^2\ell^2p^4.
\end{equation}
Neglecting negative mass states as nonphysical, Sidharth established a new fact. Namely, as the result of application of the Dirac-like linearization procedure within the modified equivalence principle (\ref{ss}) one concludes the appropriate Dirac Hamiltonian constraint which, however, differs from the standard one by an additional $\gamma^5$-term, that is proportional to 2nd power of the spatial momentum of a relativistic particle and to a minimal scale $\ell$ \cite{bgs2005ijmp1}
\begin{equation}\label{ds}
  \gamma^\mu p_\mu+mc^2+\sqrt{\alpha}\dfrac{c}{\hbar}\ell\gamma^5 p^2=0.
\end{equation}

The modified Dirac Hamiltonian constraint (\ref{ds}) formally can be deduced from the equation (\ref{ss}) rewritten in the following compact form
\begin{equation}\label{ss1}
  (\gamma^\mu p_{\mu})^2=m^2c^4+\alpha\left(\dfrac{c}{\hbar}\right)^2\ell^2p^4,
\end{equation}
where $p_\mu$ is a relativistic momentum four-vector
\begin{equation}
  p_{\mu}=\left[\begin{array}{c}E\\-cp\end{array}\right].
\end{equation}
However, in both papers as well as books Sidharth is not noticing that from the Hamiltonian constraint (\ref{ss1}) there is arising a one more additional possibility physically nonequivalent to (\ref{ds}), namely, it is given by the Dirac constraint with the correction possessing a negative sign
\begin{equation}\label{ds1}
  \gamma^\mu p_\mu+mc^2-\sqrt{\alpha}\dfrac{c}{\hbar}\ell\gamma^5 p^2=0.
\end{equation}
However, the possible physical results following from the Hamiltonian constraint (\ref{ds1}) can be deduced by application of the mirror reflection $\ell\rightarrow -\ell$ within the results following from the Dirac Hamiltonian constraint with the positive $\gamma^5$-term (\ref{ds}). We are not going to neglect also the negative mass states as nonphysical, because this situation is in strict correspondence with results obtained from the equation (\ref{ds}) by a mirror reflection in mass of a relativistic particle $m\rightarrow-m$. It means that after employing the canonical quantization in the momentum space of a relativistic particle
\begin{equation}
  E\rightarrow\hat{E}=i\hbar\partial_0\quad,\quad p\rightarrow\hat{p}=i\hbar\partial_i\quad,
\end{equation}
in general one can consider the generalized modification of Dirac equation of the form
\begin{equation}\label{ds2}
  \left(\gamma^\mu p_\mu\pm mc^2\pm\sqrt{\alpha}\dfrac{c}{\hbar}\ell\gamma^5 p^2\right)\psi=0,
\end{equation}
which describes 4 physically nonequivalent situations. Here is assumed that in analogy to the conventional Dirac theory, a solution $\psi$ of the equation (\ref{dsi}) is four component spinor
\begin{equation}
  \psi=\left[\begin{array}{c}\phi_0\\ \phi_1\\ \phi_2\\ \phi_3\end{array}\right],
\end{equation}
and that the four-dimensional Clifford algebra of the Dirac $\gamma$-matrices is given in the standard representation
\begin{eqnarray}
  \gamma^0&=&\left[\begin{array}{cc}0&\mathbf{1}_2\\\mathbf{1}_2&0\end{array}\right]\quad,\quad\gamma^i=\left[\begin{array}{cc}0&\sigma^i\\-\sigma^i&0\end{array}\right]\quad,\\
  \gamma^5&=&\gamma^0\gamma^1\gamma^2\gamma^3=i\left[\begin{array}{cc}\mathbf{1}_2&0\\0&-\mathbf{1}_2\end{array}\right]\quad,\quad \left(\gamma^5\right)^2=-\mathbf{1}_4,
\end{eqnarray}
where $\sigma$'s are the Pauli matrices
\begin{equation}
  \sigma^1=\left[\begin{array}{cc}0&1\\1&0\end{array}\right]\quad,\quad\sigma^2=\left[\begin{array}{cc}0&-i\\i&0\end{array}\right]\quad,\quad\sigma^3=\left[\begin{array}{cc}1&0\\0&-1\end{array}\right].
\end{equation}
A presence of the Dirac's matrix $\gamma^5$ in the Dirac equation (\ref{ds2}) causes that it violates parity symmetry manifestly, so in fact there is CP violation and the $\gamma^5$-term breaks the full Lorentz symmetry. For simplicity, however, it is useful to consider one of the four situations describing by the equation (\ref{ds2}), that is given by the Dirac equation modified due to the Sidharth term
\begin{equation}
  \left(\gamma^\mu \hat{p}_\mu+mc^2+\sqrt{\alpha}\dfrac{c}{\hbar}\ell\gamma^5 \hat{p}^2\right)\psi=0,\label{dsi}
\end{equation}
and finally discuss results of application of the mentioned mirror transformations.

Recently it was shown \cite{glinka} that there are some nonequivalent possibilities for establishment of the Hamiltonian from the constraint (\ref{ss}), and it crucially depends on the functional relation between a mass of a relativistic particle and a minimal scale $m(\ell)$. It leads to some nontrivial classical solutions and associated with them nonequivalent quantum theories. This energy-momentum relation is currently under astrophysics' interesting \cite{maccione}. Originally the equation (\ref{dsi}) was proposed some time ago \cite{bgs2005ijmp1} as an idea for ultra-high energy physics, but any concrete physical predictions arising from this idea still are not well-established. Currently there are only speculations possessing laconic character that the extra term violating the Lorentz symmetry manifestly lies in the new foundations of physics \cite{sidpc}. In fact its meaning is still a great riddle to the same degree as it is an amazing hope. The best test for checking the corrected theory (\ref{dsi}) and in general all the theories given by (\ref{ds2}) seem to be astrophysical phenomena \emph{i.e.} ultra-high-energy cosmic rays coming from gamma bursts sources, neutrinos coming from supernovas, and others observed in this energy region. This cognitive aspect of the thing is the motivation for reconsidering the equation (\ref{dsi}) arising due to the Snyder noncommutative geometry (\ref{nd}), and try pull out extension of well-grounded physical knowledge.

\section{Massive neutrinos}
Let us reconsider the modified Dirac equation (\ref{dsi}). In fact the Sidharth $\gamma^5$-term is the additional effect -- the shift of the conventional Dirac theory -- arising due to the Snyder noncommutative geometry of phase space $(p,x)$ of a relativistic particle (\ref{nd}). However, it does not mean that Special Relativity will be also modified - the Minkowski hyperbolic geometry of the relativistic momentum space as well as the structure of space-time in fact are preserved. The Einstein theory describes dynamics of a relativistic particle while in the philosophical as well as physical foundations of the algebra deformation we have not any arguments following from dynamics of a particle -- strictly speaking the correction is due to finite sizes of a particle. In this manner, the best interpretation of the deformation (\ref{ss}), as well as the appropriate constraint (\ref{ds}), is the energetic constraint corrected by the non-dynamical term. By this reason we propose here to take into account the formalism of the Minkowski geometry of the momentum space independently from a presence of the $\gamma^5$-term, and apply it within both the modified Einstein Hamiltonian constraint as well the modified Dirac equation.

Application of the standard identity holding in the momentum space of a relativistic particle
\begin{equation}\label{ein}
p_\mu p^\mu = \left(\gamma^\mu p_\mu\right)^2 = E^2-c^2p^2,
\end{equation}
to the modified Dirac equation (\ref{dsi}) yields the equation
\begin{equation}\label{hc}
\left[\gamma^\mu \hat{p}_\mu+mc^2+\dfrac{\sqrt{\alpha}}{\hbar c}\ell\gamma^5 \left[E^2-\left(\gamma^\mu \hat{p}_\mu\right)^2\right]\right]\psi=0,
\end{equation}
which can be rewritten as
\begin{equation}
  \left[-\dfrac{\sqrt{\alpha}}{\hbar c}\ell\gamma^5\left(\gamma^\mu \hat{p}_\mu\right)^2+\gamma^\mu \hat{p}_\mu+mc^2+\dfrac{\sqrt{\alpha}}{\hbar c}\ell\gamma^5E^2\right]\psi=0,
\end{equation}
or equivalently by using of the combination $\gamma^5\gamma^\mu p_\mu$
\begin{equation}\label{qe}
 \left[ \left(\gamma^5\gamma^\mu \hat{p}_\mu\right)^2-\epsilon\left(\gamma^5\gamma^\mu \hat{p}_\mu\right)+E^2-\epsilon mc^2\gamma^5\right]\psi=0,
\end{equation}
where $\epsilon$ is the energy
\begin{equation}
  \epsilon = \dfrac{\hbar c}{\sqrt{\alpha}\ell}.\label{eps}
\end{equation}
Note that for the Planck scale holds $\ell=\ell_{Pl}=\sqrt{\strut{\dfrac{\hbar c}{G}}}$ and the energy (\ref{eps}) coincides with the Planck energy scaled by the factor $\dfrac{1}{\sqrt{\alpha}}$
\begin{equation}
  \epsilon = \epsilon_{Pl}=\dfrac{1}{\sqrt{\alpha}}\sqrt{\strut{\dfrac{\hbar c^5}{G}}}=\dfrac{1}{\sqrt{\alpha}}M_{Pl}c^2.
\end{equation}
Similarly for the Compton scale $\ell=\ell_C=2\pi\dfrac{\hbar}{m_p c}$ is the Compton wavelength of a particle possessing the rest mass $m_p$. In this case the energy $\epsilon$ is a particle's rest energy scaled by the factor $\dfrac{1}{2\pi\sqrt{\alpha}}$
\begin{equation}
  \epsilon=\epsilon_C=\dfrac{1}{2\pi\sqrt{\alpha}}m_p c^2.
\end{equation}
If the particle has the rest mass that equals the Planck mass $m_p\equiv M_{Pl}$ then
\begin{equation}
 \ell_C=\dfrac{2\pi G}{c^2} M_{Pl}\quad,\quad\epsilon_C=\dfrac{\epsilon_{Pl}}{2\pi}.
\end{equation}
In the other words for this case the doubled Compton scale is a circumference of a circle with a radius of the Schwarzschild radius of the Planck mass (Cf. also Ref. \cite{kiefer})
\begin{equation}
  2\ell_C = 2\pi r_S\left(M_{Pl}\right)\quad,\quad r_S(m)=\dfrac{2Gm}{c^2}.
\end{equation}

The equation (\ref{qe}) expresses acting of the operator
\begin{equation}\label{oper}
  \left(\gamma^5\gamma^\mu \hat{p}_\mu\right)^2-\epsilon\left(\gamma^5\gamma^\mu \hat{p}_\mu\right)+E^2-\epsilon mc^2\gamma^5,
\end{equation}
on the Dirac spinor $\psi$. With using of elementary algebraic manipulations, however, one one can easily deduce that in fact the operator (\ref{oper}) can be rewritten in the reduced form
\begin{equation}\label{sol}
  (\gamma^5\gamma^\mu \hat{p}_\mu-\mu_+)(\gamma^5\gamma^\mu \hat{p}_\mu-\mu_-),
\end{equation}
where $M_{pm}$ are the manifestly nonhermitian quantities
\begin{equation}\label{mpm}
  \mu_{\pm}=\dfrac{\epsilon}{2}\left(1\pm\sqrt{\strut{1-\dfrac{4E^2}{\epsilon^2}}}\sqrt{\strut{1+\dfrac{4\epsilon mc^2}{\epsilon^2-4E^2}\gamma^5}}\right).
\end{equation}
Principally the quantities (\ref{mpm}) are due to the order reduction, and also cause the Dirac-like linearization.

Treating energy $E$, mass $m$, and $\epsilon$ (or equivalently the scale $\ell$) in (\ref{mpm}) as free parameters one obtains easily that formally the modified Dirac equation (\ref{dsi}) and also the generalized equation (\ref{ds2}) are equivalent to the following two nonequivalent Dirac equations
\begin{eqnarray}
  \left(\gamma^\mu \hat{p}_\mu-M_+c^2\right)\psi=0\qquad,\qquad \left(\gamma^\mu \hat{p}_\mu-M_-c^2\right)\psi=0,\label{rhc}
\end{eqnarray}
where $M_{\pm}$ are the mass matrices of the Dirac theories generated as the result of the dimensional reduction
\begin{equation}\label{mm}
  M_{\pm}=\dfrac{\epsilon}{2c^2}\left(-1\mp\sqrt{\strut{1-\dfrac{4E^2}{\epsilon^2}+\dfrac{4mc^2}{\epsilon}\gamma^5}}\right)\gamma^5.
\end{equation}
This is nontrivial result -- we have obtained usual Dirac theories, where the mass matrices $M_{pm}$ are manifestly nonhermitian $M_{\pm}^\dag\neq M_{\pm}$. However, the total effect from a minimal scale $\ell$ sits within the matrices $M_{pm}$ only, while the four-momentum operator $\hat{p}_\mu$ remains exactly the same as in both the conventional Einstein and Dirac theories. Note that this procedure formally is not incorrect - we preserve the Minkowski geometry formalism for the square of spatial momentum that in fact is the fundament of the $\gamma^5$-correction, but was not noticed or was omitted in Sidharth's papers and books. In this manner we have constructed new type mass generation mechanism which deduction within the usual frames of Special Relativity only, \emph{i.e.} for the case of vanishing sizes of the particle $\ell=0$ or equivalently for the maximal energy $\epsilon=\infty$, can not be done. Strictly speaking this mass generation mechanism is due to the order reduction in the operator (\ref{oper}) of the modified Dirac equation. However, both the mass matrices (\ref{mm}) are builded by a square root of the expression containing the matrix $\gamma^5$. Let us present now the mass matrices in equivalent way, where the Dirac matrix $\gamma^5$ will present in a linear way.

Let us see details of the mass matrices $M_{\pm}$. Fist, by application of the Taylor series expansion to the square root present in the defining formula (\ref{mm}) one obtains
\begin{eqnarray}
\sqrt{\strut{1-\dfrac{4E^2}{\epsilon^2}+\dfrac{4mc^2}{\epsilon}\gamma^5}}&=&\sqrt{\strut{1-\dfrac{4E^2}{\epsilon^2}}}\sqrt{\strut{1+\dfrac{\dfrac{4mc^2}{\epsilon}}{1-\dfrac{4E^2}{\epsilon^2}}\gamma^5}}=\nonumber\\
&=&\sqrt{\strut{1-\dfrac{4E^2}{\epsilon^2}}}\sum_{n=0}^\infty \binom{1/2}{n}\left(\dfrac{\dfrac{4mc^2}{\epsilon}}{1-\dfrac{4E^2}{\epsilon^2}}\gamma^5\right)^n,\label{tay}
\end{eqnarray}
where the following notation was used $$\binom{n}{k}=\dfrac{\Gamma(n+1)}{\Gamma(k+1)\Gamma(n+1-k)}$$ that is the generalized Newton binomial symbol. Employing now the $\gamma^5$-matrix properties -- \emph{i.e.} $\left(\gamma^5\right)^{2n}=-1$, and $\left(\gamma^5\right)^{2n+1}=-\gamma^5$ -- one decompose the sum present in the last term of (\ref{tay}) onto the two component
\begin{eqnarray}
  \!\!\!\!\!\!\!\!\!\!\!\!\!\!\!\!\!\!\!\!&&\sum_{n=0}^\infty \binom{1/2}{n}\left(\dfrac{\dfrac{4mc^2}{\epsilon}}{1-\dfrac{4E^2}{\epsilon^2}}\gamma^5\right)^n=\nonumber\\
  \!\!\!\!\!\!\!\!\!\!\!\!\!\!\!\!\!\!\!\!&&=-\sum_{n=0}^\infty \binom{1/2}{2n}\left(\dfrac{\dfrac{4mc^2}{\epsilon}}{1-\dfrac{4E^2}{\epsilon^2}}\right)^{2n}-\sum_{n=0}^\infty \binom{1/2}{2n+1}\left(\dfrac{\dfrac{4mc^2}{\epsilon}}{1-\dfrac{4E^2}{\epsilon^2}}\right)^{2n+1}\gamma^5.\label{comp}
\end{eqnarray}
Direct application of standard summation procedure allows to establish the sums presented in the both components in (\ref{comp}) in a compact form
\begin{eqnarray}
  \!\!\!\!\!\!\!\!\!\!\!\!\!\!\!\!\!\!\!\!&&\sum_{n=0}^\infty\binom{1/2}{2n}\left(\dfrac{\dfrac{4mc^2}{\epsilon}}{1-\dfrac{4E^2}{\epsilon^2}}\right)^{\!\!\!2n}=\sqrt{\strut{1+\dfrac{\dfrac{4mc^2}{\epsilon}}{1-\dfrac{4E^2}{\epsilon^2}}}}+\sqrt{\strut{1-\dfrac{\dfrac{4mc^2}{\epsilon}}{1-\dfrac{4E^2}{\epsilon^2}}}},\vspace*{10pt}\\
  \!\!\!\!\!\!\!\!\!\!\!\!\!\!\!\!\!\!\!\!&&\sum_{n=0}^\infty\binom{1/2}{2n+1}\left(\dfrac{\dfrac{4mc^2}{\epsilon}}{1-\dfrac{4E^2}{\epsilon^2}}\right)^{\!\!\!2n+1}=\sqrt{\strut{1+\dfrac{\dfrac{4mc^2}{\epsilon}}{1-\dfrac{4E^2}{\epsilon^2}}}}-\sqrt{\strut{1-\dfrac{\dfrac{4mc^2}{\epsilon}}{1-\dfrac{4E^2}{\epsilon^2}}}}.
\end{eqnarray}
In this manner finally one sees easily that both the mass matrices $M_{\pm}$ possess the following formal decomposition
\begin{equation}\label{mm1}
  M_{\pm}=\mathfrak{H}(M_{\pm})+\mathfrak{A}(M_{\pm}),
\end{equation}
where $\mathfrak{H}(M_{\pm})$ is hermitian part of $M_{\pm}$
\begin{equation}
  \mathfrak{H}(M_{\pm})=\pm\dfrac{\epsilon}{2c^2}\left[\sqrt{\strut{1-\dfrac{4E^2}{\epsilon^2}}}\left(\sqrt{\strut{1+\dfrac{\dfrac{4mc^2}{\epsilon}}{1-\dfrac{4E^2}{\epsilon^2}}}}-\sqrt{\strut{1-\dfrac{\dfrac{4mc^2}{\epsilon}}{1-\dfrac{4E^2}{\epsilon^2}}}}\right)\right],
\end{equation}
and $\mathfrak{A}(M_{\pm})$ is antihermitian part of $M_{\pm}$
\begin{equation}
  \mathfrak{A}(M_{\pm})=-\dfrac{\epsilon}{2c^2}\left[1\pm\sqrt{\strut{1-\dfrac{4E^2}{\epsilon^2}}}\left(\sqrt{\strut{1+\dfrac{\dfrac{4mc^2}{\epsilon}}{1-\dfrac{4E^2}{\epsilon^2}}}}+\sqrt{\strut{1-\dfrac{\dfrac{4mc^2}{\epsilon}}{1-\dfrac{4E^2}{\epsilon^2}}}}\right)\right]\gamma^5.
\end{equation}

By application of elementary algebraic manipulations one sees that equivalently the mass matrices $M_{\pm}$ can be decomposed into the basis of the commutating projectors $\left\{\Pi_i:\dfrac{1+\gamma^5}{2},\dfrac{1-\gamma^5}{2}\right\}$,
\begin{equation}\label{mm2}
  M_{\pm}=\sum_i\mu_i^\pm\Pi_i=\mu_R^\pm\dfrac{1+\gamma^5}{2}+\mu_L^\pm\dfrac{1-\gamma^5}{2},
\end{equation}
where
\begin{eqnarray}
  \mu_R^\pm&=&-\dfrac{1}{c^2}\left(\dfrac{\epsilon}{2}\pm\sqrt{\strut{\epsilon^2-4\epsilon mc^2-4E^2}}\right),\label{mu1}\\
  \mu_L^\pm&=&\dfrac{1}{c^2}\left(\dfrac{\epsilon}{2}\pm\sqrt{\strut{\epsilon^2+4\epsilon mc^2-4E^2}}\right),\label{mu2}
\end{eqnarray}
are projected masses related to the theories with signs $\pm$ in the matrix mass. By application of the obvious relations for the projectors $\Pi_i^\dag\Pi_i=\mathbf{1}_4$, $\Pi_1\Pi_2=\dfrac{1}{2}\mathbf{1}_4$, $\Pi_1^\dag=\Pi_2$ and $\Pi_1+\Pi_2=\mathbf{1}_4$ one obtains
\begin{equation}
  M_\pm M_\pm^\dag=\dfrac{(\mu_R^\pm)^2+(\mu_L^\pm)^2}{2}\mathbf{1}_4.
\end{equation}
Introducing the right- and left-handed chiral Weyl fields
\begin{equation}\label{rl}
\psi_R=\dfrac{1+\gamma^5}{2}\psi\quad,\quad\psi_L=\dfrac{1-\gamma^5}{2}\psi,
\end{equation}
where the Dirac spinor $\psi$ is a solution of the appropriate Dirac equations (\ref{rhc}), both the theories (\ref{rhc}) can be rewritten as the system of two equations
\begin{equation}\label{neu}
  \left(\gamma^\mu \hat{p}_\mu+\mu^+ c^2\right)\left[\begin{array}{c}\psi_R^+\\ \psi_L^+\end{array}\right]=0\qquad,\qquad \left(\gamma^\mu \hat{p}_\mu+\mu^- c^2\right)\left[\begin{array}{c}\psi_R^-\\ \psi_L^-\end{array}\right]=0,
\end{equation}
where the mass matrices $\mu^\pm$ are hermitian now
\begin{equation}\label{mm3}
  \mu^\pm=\left[\begin{array}{cc}\mu_R^\pm&0\\0&\mu_L^\pm\end{array}\right]=\left[\begin{array}{cc}\mu_R^\pm&0\\0&\mu_L^\pm\end{array}\right]^\dag,
\end{equation}
and $\psi_{R,L}^\pm$ are the chiral fields related to the mass matrices $\mu_{\pm}$ respectively. Note that the masses (\ref{mu1}) and (\ref{mu2}) are invariant with respect to choice of the Dirac matrices $\gamma^\mu$ representation. By this way they have physical character. It is interesting that for the mirror reflection in a minimal scale $\ell\rightarrow-\ell$ (or equivalently for the change $\epsilon\rightarrow-\epsilon$) we have the exchange $\mu_R^\pm\leftrightarrow\mu_L^\pm$ while the chiral Weyl fields are the same. In the case of the mirror reflection in the original mass $m\rightarrow-m$ one has the exchange $\mu_R^\pm\leftrightarrow-\mu_L^\pm$. The case of originally massless states $m=0$ is also intriguing from theoretical point of view. From the formulas (\ref{mu1}) and (\ref{mu2}) one sees easily that in this case $\mu_R=-\mu_L$. In the case of generic Einstein theory $\ell=0$ one has
\begin{equation}
\mu_{R}^\pm=\left\{\begin{array}{cc}-\infty&~\mathrm{for}~+\\ \infty&~\mathrm{for}~-\end{array}\right.\qquad,\qquad \mu_{L}^\pm=\left\{\begin{array}{cc}\infty&~\mathrm{for}~+\\ -\infty&~\mathrm{for}~-\end{array}\right. .
\end{equation}
In general, however, for formal correctness of the projection splitting (\ref{mm2}) both the neutrinos masses (\ref{mu1}) and (\ref{mu2}) must be real numbers; strictly speaking when the masses are complex numbers the decomposition (\ref{mm2}) does not yield hermitian mass (\ref{mm3}, so that the presented construction does not hold and by this reason must be replaced by other one.

In the conventional Weyl theory approach neutrinos are massless. In this manner it is evident that employing the Snyder noncommutative geometry generates a new obvious nontriviality -- \emph{the kinetic mass generation mechanism that leads to the theory of massive neutrinos}. It must be emphasized that in Sidharth's books and papers a possibility of neutrino masses was only laconically mentioning as ''due to mass term'', where by the mass term the author understands the $\gamma^5$-term in the modified Dirac equation (\ref{dsi}). In fact it is not mass term in the common sense of the Standard Model being currently the theory of elementary particles and fundamental interactions. Strictly speaking Sidharth's statements are incorrect, because we have generated the massive neutrinos due to two-step mechanism - the first was the order reduction of the modified Dirac equation (\ref{dsi}), and the second was decomposition of the received mass matrices (\ref{mm}) into the projectors basis and introducing the chiral Weyl fields in the usual way (\ref{rl}). It must be emphasized that a mass generation mechanism is manifestly absent in Sidharth's contributions and the line of thinking presented there is completely different, omits many interesting physical and mathematical details, and in general does not look like constructive (Cf. \emph{e.g.} Ref. \cite{bgs2005ijmp2}). However, in the result of the procedure proposed above, \emph{i.e.} by application of the Dirac equation with the $\gamma^5$-term (\ref{ds}) and direct preservation within this equation the Einstein--Minkowski relativity (\ref{ein}), we have generated the system of equations (\ref{neu}) which describes two left- $\psi_L^\pm$ and two right- $\psi_R^\pm$ chiral massive Weyl fields, \emph{i.e.} we have yielded massive neutrinos, related to both cases - any originally massive $m\neq0$ as well as for originally massless $m=0$ states. By this reason in the proposed approach the notion \emph{neutrino} takes an essentially new physical meaning; it is a chiral field due to any massive and massless quantum state. Moreover, we have obtained the two massive Weyl theories (\ref{neu}), so that totally with a one quantum state there are associated 4 massive neutrinos.

\section{The chiral condensate}
Let us notice that if we want to construct the Lorentz invariant Lagrangian  $\mathcal{L}$ of the gauge field theory characterized by the Euler--Lagrange equations of motion (\ref{neu}) for both massive Weyl theories we should put
\begin{eqnarray}\label{lag}
  \mathcal{L}^\pm=\bar{\psi}_R^\pm\gamma^\mu \hat{p}_\mu\psi_R^\pm + \bar{\psi}_L^\pm\gamma^\mu \hat{p}_\mu\psi_L^\pm+\mu_R^\pm c^2\bar{\psi}_R^\pm\psi_R^\pm+\mu_L^\pm c^2\bar{\psi}_L^\pm\psi_L^\pm,
\end{eqnarray}
where $\bar{\psi}_{R,L}^\pm=\left(\psi_{R,L}^\pm\right)^\dag\gamma^0$ are the Dirac adjoint of $\psi_{R,L}^\pm$, and take into considerations rather the sum of both partial gauge field theories (\ref{lag})
\begin{equation}\label{tlag}
\mathcal{L}=\mathcal{L}^+ +\mathcal{L}^-,
\end{equation}
as the Lagrangian of the appropriate full gauge field theory. One can see straightforwardly that the both partial gauge field theories (\ref{lag}) exhibit the (local) chiral symmetry $SU(2)_R^\pm\otimes SU(2)_L^\pm$
\begin{equation}
  \left\{\begin{array}{c}\psi_R^\pm\rightarrow \exp \left\{i\theta_R^\pm\right\}\psi_R^\pm\\
  \psi_L^\pm\rightarrow\psi_L^\pm\end{array}\right. \quad\mathrm{or}\quad \left\{\begin{array}{c}\psi_R^\pm\rightarrow\psi_R^\pm\\
  \psi_L^\pm\rightarrow \exp\left\{i\theta_L^\pm\right\}\psi_L^\pm\end{array}\right. ,
\end{equation}
the vector symmetry $U(1)_V^\pm$
\begin{equation}
  \left\{\begin{array}{c}\psi_R^\pm\rightarrow \exp\left\{i\theta^\pm\right\}\psi_R^\pm\\
  \psi_L^\pm\rightarrow \exp\left\{i\theta^\pm\right\}\psi_L^\pm\end{array}\right. ,
\end{equation}
and the axial symmetry $U(1)_A^\pm$
\begin{equation}
  \left\{\begin{array}{c}\psi_R^\pm\rightarrow \exp\left\{-i\theta^\pm\right\}\psi_R^\pm\\
  \psi_L^\pm\rightarrow \exp\left\{i\theta^\pm\right\}\psi_L^\pm\end{array}\right. .
\end{equation}
In this manner the total symmetry group is the composed $SU(3)_C^{TOT}$
\begin{equation}
  SU(3)_C^{TOT}=SU(3)_C^+\oplus SU(3)_C^-,
\end{equation}
where $SU(3)_C^\pm$ are the global (chiral) 3-flavor gauge symmetries related to each of the gauge theories (\ref{lag}), \emph{i.e.}
\begin{eqnarray}\label{gs}
\!\!\!\!\!\!\!\!\!\!&SU(2)_R^+\otimes SU(2)_L^+\otimes U(1)_V^+\otimes U(1)_A^+\equiv SU(3)^+\otimes SU(3)^+=SU(3)_C^+,&\\
\!\!\!\!\!\!\!\!\!\!&SU(2)_R^-\otimes SU(2)_L^-\otimes U(1)_V^-\otimes U(1)_A^-\equiv SU(3)^-\otimes SU(3)^-=SU(3)_C^-,&
\end{eqnarray}
describing 2-flavor massive free quarks -- \emph{the neutrinos} in our proposition. However, by using of the relations for the Weyl fields (\ref{rl}) and applying algebraic manipulations of the Dirac $\gamma$-algebra (as \emph{e.g.} $\left\{\gamma^\mu,\gamma^5\right\}=0$) one has
\begin{eqnarray}
  \left(1\mp\gamma^5\right)\gamma^0\left(1\pm\gamma^5\right)&=&\pm2\gamma^0\gamma^5,\\
  \left(1\mp\gamma^5\right)\gamma^0\gamma^\mu\left(1\pm\gamma^5\right)&=&2\gamma^0\gamma^5,
\end{eqnarray}
and hence contribution to the right hand side of (\ref{lag}) are
\begin{eqnarray}
\bar{\psi}_{R,L}^\pm\gamma^\mu p_\mu \psi_{R,L}^\pm&=&\dfrac{1}{2}\bar{\psi^\pm}\gamma^\mu p_\mu\psi^\pm,\label{d1}\\
\mu_{R,L}^\pm c^2\bar{\psi}^\pm_{R,L}\psi_{R,L}^\pm&=&\pm \dfrac{\mu_{R,L}^\pm}{2}c^2\bar{\psi^\pm}\gamma^5\psi^\pm,\label{d2}
\end{eqnarray}
where $\bar{\psi^\pm}=\left(\psi^\pm\right)^\dag\gamma^0$ is the Dirac adjoint of the Dirac fields $\psi^\pm$ related to the Weyl chiral fields by the transformation (\ref{rl}). Both (\ref{d1}) and (\ref{d2}) are the Lorentz invariants. In result the global chiral Lagrangian (\ref{tlag}) can be elementary lead to the following form
\begin{eqnarray}
  \mathcal{L}&=&\bar{\psi^+}\left(\gamma^\mu \hat{p}_\mu+\mu_{eff}^+ c^2\right)\psi^++\bar{\psi^-}\left(\gamma^\mu \hat{p}_\mu+\mu_{eff}^-c^2\right)\psi^-=\label{tlag1a}\\
  &=&\bar{\Psi}\left(\gamma^\mu\hat{p}_{\mu}+M_{eff}c^2\right)\Psi,\label{tlag1b}
\end{eqnarray}
where $\mu_{eff}^{\pm}$ are the effective mass matrices of the gauge fields $\psi^\pm$, and $M_{eff}$ is the mass matrix of the effective composed field $\Psi=\left[\begin{array}{c}{\psi^+}\\ {\psi^-}\end{array}\right]$
\begin{eqnarray}
  \mu_{eff}^\pm&=&\dfrac{\mu_R^\pm-\mu_L^\pm}{2}\gamma^5,\\
  M_{eff}&=&\left[\begin{array}{cc}{\mu^+_{eff}}&0\\0&{\mu^-_{eff}}\end{array}\right].
\end{eqnarray}
Both the mass matrices $\mu^\pm_{eff}$ are hermitian or antihermitian -- it depends on a choice of representation, so the same property has the mass matrix $M_{eff}$. Obviously, the full gauge field theory (\ref{tlag1a}), or equivalently (\ref{tlag1b}), is invariant with respect to the composed gauge symmetry $SU(2)_V^{TOT}$ transformation
\begin{equation}
SU(2)_V^{TOT}=SU(2)_V^+\oplus SU(2)_V^-,
\end{equation}
where $SU(2)_V^\pm$ are the $SU(2)\otimes SU(2)$ transformations used to each of the gauge fields $\psi^\pm$
\begin{equation}
  \left\{\begin{array}{c}\psi^\pm\rightarrow \exp\left\{i\theta^\pm\right\}\psi^\pm\\ \bar{\psi^\pm}\rightarrow \bar{\psi^\pm}\exp\left\{-i\theta^\pm\right\}\end{array}\right..
\end{equation}
This means that for the full gauge field theory the composed global chiral symmetry $SU(3)_C^{TOT}$ is spontaneously broken to its subgroup -- the composed isospin group $SU(2)_V^{TOT}$
\begin{equation}
SU(3)_C^{TOT}\longrightarrow SU(2)_V^{TOT}.
\end{equation}
Physically it should be interpreted as the symptom of an existence of the chiral condensate of massive neutrinos being a composition of two chiral condensates, that is the composed effective field theory $SU(2)_V^{TOT}=(SU(2)^+\otimes SU(2)^+)\oplus(SU(2)^-\otimes SU(2)^-)$ \cite{weinberg}. However, by the composed global chiral gauge symmetry $SU(3)_C^{TOT}$, the gauge theory (\ref{lag}) looks like formally as the theory of free massive quarks which do not interact; this is the situation similar to  Quantum Chromodynamics \cite{greiner}, but in the studied case we have formally a composition of two QCDs. For each of the QCDs the space of fields is different then in usual QCD - there are two massive chiral fields only -- the left- and right-handed Weyl fields, thats are the massive neutrinos by our proposition. The chiral condensate of massive neutrinos (\ref{tlag1b}) is the result beyond the Standard Model, but essentially it can be included into the theory as the new contribution.

\section{Discussion}
It must be emphasized that the energy-momentum relation (\ref{ss}) modified due to the Snyder model of noncommutative geometry (\ref{nd}) differs from the usual Special Relativity's relation. In particular as is self-evident from the Hamiltonian constraint (\ref{ss}), there is an extra contribution to the Einstein special equivalence principle due to the additional $\ell^2$-term. This is brought out very clearly in the manifestly nonhermitian Dirac equations (\ref{rhc}), as well as in the hermitian massive Weyl equations (\ref{neu}) describing the neutrinos in our proposition. A massless neutrino in the conventional Weyl theory is now seen to argue as mass, and further, this mass has a two left component and a two right component, as show in (\ref{mm1}) and (\ref{mm2}). Once this is recognized, the mass matrix which otherwise appears nonhermitian, turs out to be actually hermitian, as seen in (\ref{mm3}), but if and only if when the masses (\ref{mm1}) and (\ref{mm2}) are real. There is no any restrictions, however, for their sign - the masses can be positive as well as negative. In other words the underlying Snyder noncommutative geometry (\ref{nd}) is reflected in the modified energy-momentum relation (\ref{ds}) naturally gives rise to the mass of the neutrino. It was laconically suggested as a possible result in the Ref. \cite{bgs2005ijmp2}, however, with no any concrete calculations and proposals for a mass generation mechanism. It must be remembered that in the Standard Model the neutrino is massless, but the Super--Kamiokande experiments in the late nineties showed that the neutrino does indeed have a mass and this is the leading motivation to an exploration of models beyond the Standard Model, as the model presented in this paper. In this connection it is also relevant to mention that currently the Standard Model requires the Higgs Mechanism for the generation of mass in general, though the Higgs particle has been undetected for forty five years and it is hoped will be detected by the Large Hadron Collider, after it is recommissioned. We hope for next development of the both presented massive neutrino model, as well as the chiral condensate.

\section*{Acknowledgements}
Author benefitted many valuable discussions from Profs. B. M. Barbashov, V. N. Pervushin and A. B. Arbuzov. Interaction with Prof. A. P. Isaev in 2007 was fruitful. Communication with Dr. B. G. Sidharth was enlightening. Remarks of the referees were very helpful for corrections of the primary notes.

\end{document}